\documentclass[11pt]{article}

\usepackage{latexsym}

\title{The Complexity of Poor Man's Logic\thanks{
Corrected version of~\cite{hem01}. The changes are
in Section~\ref{s:other}, where Theorem~\ref{conp}(2)
was added to handle two missing cases.}}

\author{Edith Hemaspaandra\thanks{
Supported in part 
by grants NSF-INT-9815095 and NSF-CCR-0311021.
Work done in part while visiting the University
of Amsterdam. Email: {\tt eh@cs.rit.edu}.  }\\
Department of Computer Science\\
Rochester Institute of Technology\\ 
102 Lomb Memorial Drive\\ 
Rochester, NY 14623-5608}

\date{}

\newenvironment{proof}{\noindent {\bf Proof.}\quad}{\qed}
\newtheorem{theorem}{Theorem}[section]

\newtheorem{corollary}[theorem]{Corollary}

\def\qed{{\ \nolinebreak\hfill\mbox{$\Box$}}}	
\newcommand{\pair}[1]{\mbox{$<\!#1\!>$}}
\newcommand{\K}{\mbox{K}}
\newcommand{\Kleqone}{{\cal F}_{\leq 1}}
\newcommand{\Kleqtwo}{{\cal F}_{\leq 2}}
\newcommand{\Kgeqone}{{\cal F}_{\geq 1}}

\newcommand{\Sfive}{\mathrm{S5}}
\newcommand{\mbar}{\ |\ }
\newcommand{\md}{\mbox{\it md}}
\newcommand{\smd}{\mbox{\footnotesize{\it md}}}
\newcommand{\ALEN}{${\cal ALEN}$}
\newcommand{\ALE}{${\cal ALE}$}
\newcommand{\calL}{{\cal L}}
\newcommand{\atneg}{{}^{\mbox{\normalsize{$\overline{~}$}}}}
\newcommand{\lpml}{{\cal L}(\{\atneg, \wedge, \Box, \Diamond\})}
\newcommand{\allOp}{\{\neg, \atneg, \wedge, \vee, \Box, \Diamond, \true,
\false\}}
\newcommand{\ol}[1]{\overline{#1}}

\newcommand{\true}{\mbox{\it true}}
\newcommand{\false}{\mbox{\it false}}

\newcommand{\phiexp}{\phi_{\mbox{\it exp}}}
\newcommand{\labelfalse}{\mbox{\it label\_false}}

\newcommand{\Pow}{\mbox{\it Pow}}

\begin{document}
\maketitle

\begin{abstract}
Motivated by description logics, we investigate what
happens to the complexity of modal satisfiability problems
if we only allow formulas built from literals, $\wedge$, $\Diamond$,
and $\Box$.  Previously, the only known result was that
the complexity of the satisfiability problem for $\K$ dropped from
PSPACE-complete to coNP-complete (Schmidt-Schauss and Smolka~\cite{ss} and
Donini et al.~\cite{dhlsnn}).
In this paper we show that not all modal logics behave like $\K$.
In particular, we show that the complexity of the satisfiability
problem with respect to frames in which each world has at least
one successor drops from PSPACE-complete to P, but that in contrast
the satisfiability problem with respect to the class of frames in which
each world has at most two successors remains PSPACE-complete.
As a corollary of the latter result, we also solve the open problem
from Donini et al.'s complexity classification of
description logics~\cite{dlnn}. 
In the last section, we classify the complexity of the satisfiability
problem for $\K$ for all other restrictions on the set of operators.
\end{abstract}


\sloppy

\section{Introduction}
Since consistent normal modal logics contain propositional
logic, the satisfiability problems for all these logics are
automatically NP-hard.
In fact, as shown by Ladner~\cite{ladner}, many of them are
even PSPACE-hard.

But we don't always need all of propositional logic.
For example, in some applications we may use only a finite set
of propositional variables. Propositional satisfiability
thus restricted is in P, and, as shown by Halpern~\cite{halpern},
the complexity of satisfiability problems for some
modal logics restricted in the same way also decreases. 
For example, the complexity of $\Sfive$ satisfiability drops from
NP-complete to P. On the other hand, $\K$ satisfiability remains
PSPACE-complete.
The same restriction for linear temporal logics
was studied in  Demri and Schnoebelen~\cite{demri98}.

Restricting the number of propositional variables is not the
only propositional restriction on  modal
logics that occurs in the literature. For example,
the description logic \ALE\ can be viewed as multi-modal
$\K$ where the formulas are built from literals, $\wedge$, $\Diamond$s,
and $\Box$s. 

As in the case of a fixed number of propositional variables,
satisfiability for propositional logic
for formulas built from literals and $\wedge$
is easily seen to be in P.
After all, in that case every propositional formula 
is the conjunction of literals.  Such a formula is 
satisfiable if and only if there is no propositional
variable $p$ such that both $p$ and $\ol{p}$ are conjuncts of 
the formula.

Hence, satisfiability for modal logics for formulas
built from literals, $\wedge$, $\Box$, and $\Diamond$
is not automatically NP-hard.
Of course, it does not necessarily follow that the complexity
of modal satisfiability problems will drop significantly.
The only result that was previously known is that
the complexity of $\K$ satisfiability (i.e., satisfiability
with respect to the class of all frames) drops from
PSPACE-complete to coNP-complete. The upper bound was shown
by Schmidt-Schauss and Smolka~\cite{ss}, and the lower bound
by Donini et al.~\cite{dhlsnn}. It should be noted that these
results were shown in the context of description logics
(a.k.a.\ concept languages), so that  the 
notation in these papers is quite different from ours.\footnote{
Certain description logics can be viewed as syntactic variations
of modal logics in the following way:
the universal concept corresponds to $\true$,
the empty concept corresponds to $\false$,
atomic concepts correspond to propositional variables,
atomic negation corresponds to propositional negation,
complementation corresponds to negation,
intersection corresponds to conjunction,
union corresponds to disjunction,
universal role quantifications correspond to $\Box$ operators, and
existential role quantifications correspond to $\Diamond$
operators~\cite{schild}.}
In addition, their language contains the constants
$\true$ and $\false$.  However, it is easy to simulate
these constants by propositional variables.
See Theorem~\ref{conp}(1) for details.

In this paper we investigate if it is always the case that
the complexity of the satisfiability problem decreases if
we only look at formulas that are built from literals, $\wedge$, $\Diamond$,
and $\Box$, and if so, 
if there are upper or lower bounds on the amount that the complexity
drops.

We will show that not all logics behave like $\K$.
Far from it, by looking at simple restrictions on the number of
successors that are allowed for each world in a frame, we obtain 
different levels of complexity, making apparent a subtle interplay
between frames and operators.
In particular, we will show that
\begin{enumerate}
\item The complexity of the satisfiability problem with respect to
linear frames drops from NP-complete to P.
\item The complexity of the satisfiability problem with respect
to 
\setlength{\unitlength}{0.5pt}
\begin{picture}(30,30)(18,-10)
\put(30,10){\circle*{5}}
\put(27,6){\vector(-3,-4){9}}
\put(30,6){\vector(0,-1){12}}
\put(33,6){\vector(3,-4){9}}
\put(15,-10){\circle*{5}}
\put(30,-10){\circle*{5}}
\put(45,-10){\circle*{5}}
\end{picture}
remains NP-complete.
\item The complexity of the satisfiability problem with respect to
frames in which every world has at least one successor
drops from PSPACE-complete to P.
\item The complexity of the satisfiability problem with respect to
frames in which every world has at most two successors
remains PSPACE-complete.
\end{enumerate}
As a corollary of the last result, we also solve the open problem
from Donini et al.'s complexity classification of
description logics~\cite{dlnn}.

In the last section, we completely classify the complexity of the
satisfiability problem (with respect to the class of all frames)
for all possible restrictions on the set of operators
allowed, to gain more insight in the sources of complexity
for modal logics. It turns out that the restriction studied
in this paper, which we will call poor man's logic, is the only
(constant-free) fragment whose satisfiability problem is so unusual.
For all other restrictions, the satisfiability problems are
PSPACE-complete, NP-complete, or in P.
These are exactly the complexity classes that one would expect
to show up in this context.

\section{Definitions}

We will first briefly review syntax, Kripke semantics, and some basic
terminology for modal logic. 

\subsection*{Syntax}
The set of ${\cal L}$ formulas is inductively defined as
follows. (As usual, we assume that we have a countably infinite set
of propositional variables.)
\begin{itemize}
\item $p$ and $\ol{p}$ are ${\cal L}$ formulas for every propositional
variable $p$,
\item if $\phi$ and $\psi$ are ${\cal L}$ formulas, then so are 
$\phi \wedge \psi$ and $\phi \vee \psi$, and
\item if $\phi$ is an ${\cal L}$ formula, then $\Box \phi$  and
$\Diamond \phi$ are ${\cal L}$ formulas.
\end{itemize}
We will identify $\ol{\ol{p}}$ with $p$.

The {\em modal depth} of a formula $\phi$ (denoted by $\md(\phi)$)
is the depth of nesting of the modal operators $\Box$ and $\Diamond$.

\subsection*{Semantics}
A {\em frame} is a tuple $F = \pair{W,R}$ where $W$
is a non-empty set of possible worlds, and $R$
is a binary relation on $W$
called the accessibility relation.

A  {\em model} is of the form $M = \pair{W,R,\pi}$
such that 
$\pair{W,R}$ is a frame (we say that $M$ is {\em based
on} this frame), and $\pi$ is a function from the set of propositional
variables to $\Pow(W)$: a valuation, i.e., $\pi(p)$ is the set of worlds
in which $p$ is true. For $\phi$ an ${\cal L}$ formula, 
we will write $M,w \models \phi$ for $\phi$ is {\em true /satisfied at
$w$ in $M$}. The truth relation $\models$ is defined with induction
on $\phi$ in the following way.
\begin{itemize}
\item $M,w \models p$ iff $w \in \pi(p)$ for $p$ a propositional 
variable.
\item $M,w \models \ol{p}$ iff $w \not \in \pi(p)$ for $p$
a propositional variable.
\item $M,w \models \phi \wedge \psi$ iff $M,w \models \phi$ and $M,w
\models \psi$.
\item $M,w \models \phi \vee \psi$ iff $M,w \models \phi$ or $M,w
\models \psi$.
\item $M,w \models \Box \phi$ iff $\forall w' \in W [w R w'
\Rightarrow M,w' \models \phi]$.
\item $M,w \models \Diamond \phi$ iff $\exists w' \in W [w R w'
\mbox{ and } M,w' \models \phi]$.
\end{itemize}
The {\em size} of a model or frame 
is the number of worlds in the model or frame.

The notion of satisfiability can be extended to models, frames, and classes
of frames in the following way.
$\phi$ is satisfied in model $M$ if $M,w \models \phi$ for some world
$w$ in $M$, 
$\phi$ is satisfiable in frame $F$ ($F$ satisfiable) if 
$\phi$ is satisfied in $M$ for some model $M$ based on $F$, and
$\phi$ is satisfiable with respect to class of frames
${\cal F}$ (${\cal F}$ satisfiable) if 
$\phi$ is satisfiable in some frame $F \in {\cal F}$.

As usual, we will look at satisfiability with respect to classes
of frames. For a class of frames ${\cal F}$, the satisfiability
problem with respect to ${\cal F}$ is the problem of determining,
given an ${\cal L}$ formula $\phi$, whether $\phi$ is ${\cal F}$ satisfiable.
For a complete logic $L$, we will sometimes view $L$
as the class of frames where $L$ is valid.  For example, we will
speak of $\K$ satisfiability when we mean satisfiability with
respect to all frames. Likewise, we will on occasion identify a class
of frames with its logic, i.e., with the set of formulas
valid on this class of frames.

\subsection*{Poor Man's Logic}
The set of poor man's formulas is the set of ${\cal L}$
formulas that do not contain $\vee$.
The poor man's satisfiability problem with respect to ${\cal F}$
is the problem of determining, given a poor man's formula $\phi$,
whether $\phi$ is ${\cal F}$ satisfiable.

In poor man's language, we
will view $\wedge$ as a multi-arity
operator, and we will assume that all conjunctions are ``flattened,''
that is,
a conjunct will not be a conjunction.
Thus, a formula $\phi$ in this language is of the following form:
$\phi = \Box \psi_1 \wedge \cdots \wedge  \Box \psi_k \wedge
\Diamond \xi_1 \wedge \cdots \wedge  \Diamond \xi_m \wedge
\ell_1 \wedge \cdots \wedge \ell_s$, where
the $\ell_i$s are literals.

In all but the last section of this paper, we will 
compare the complexity of satisfiability to the complexity
of poor man's satisfiability with respect to the same class of
frames. We are interested in simple restrictions on the number
of successor worlds that are
allowed. Let $\Kleqone$, $\Kleqtwo$, and $\Kgeqone$ be the classes
of frames in which every world has at most one, at most two,
and at least one successor, respectively. (Note that 
$\Kgeqone$ corresponds to the logic KD.)

\section{Poor Man's Versions of NP-complete Satisfiability Problems}
We already know that the poor man's version of an NP-complete
modal satisfiability problem can be in P. Look for example
at satisfiability with respect to the class of frames where no world has
a successor. This is plain propositional logic in disguise, and it inherits
the complexity behavior of propositional logic.
As mentioned
in the introduction, the complexity of satisfiability
drops from NP-complete to P.

In this section, we will give an example of a non-trivial
modal logic with the same behavior.  We will show that the poor
man's version of satisfiability with respect to linear frames
is in P. In contrast,
we will also give a very simple example of a modal logic
where the complexity of poor man's satisfiability remains
NP-complete.

\begin{theorem}
\label{npp}
Satisfiability with respect to  $\Kleqone$ is NP-complete
and poor man's satisfiability with respect to $\Kleqone$ is in P.
\end{theorem}
\begin{proof}
Clearly, $\Kleqone$ satisfiability is in NP (and thus NP-complete), since 
every satisfiable formula is satisfiable on a linear frame with
$\leq \md(\phi)$ worlds, where $\md(\phi)$ is the modal depth of $\phi$. 
This immediately gives the following NP algorithm for $\Kleqone$
satisfiability:
Guess a linear frame of size $\leq \md(\phi)$, and for every world
in the frame, guess a valuation on the propositional variables that occur
in $\phi$.
Accept if and only if the guessed model satisfies $\phi$.

It is easy to see that the following polynomial-time algorithm
decides poor man's satisfiability
with respect to $\Kleqone$.
Let $\phi = \Box \psi_1 \wedge \cdots \wedge  \Box \psi_k \wedge
\Diamond \xi_1 \wedge \cdots \wedge  \Diamond \xi_m \wedge
\ell_1 \wedge \cdots \wedge \ell_s$, where the $\ell_i$s are 
literals.
$\phi$ is $\Kleqone$ satisfiable if and only if
\begin{itemize}
\item $\ell_1 \wedge \cdots \wedge \ell_s$ is satisfiable
(that is,  for all $i$ and $j$, $\ell_i \neq \ol{\ell_j}$), and
\item
\begin{itemize}
\item $m = 0$, (that is, $\phi$ does not contain conjuncts of the form
$\Diamond \xi$, in which case the formula is satisfied in a world with
no successors), or
\item $\bigwedge_{i = 1}^k \psi_i \wedge \bigwedge_{i = 1}^m \xi_i$
is $\Kleqone$ satisfiable (the world has exactly one successor).
\end{itemize}
\end{itemize}
\end{proof}

From the previous example, 
you might think that the poor man's versions of logics
with the poly-size frame property are in P, or even that the
poor man's versions of all NP-complete satisfiability
problems are in P.  Not so. The following theorem gives a very
simple counterexample.

\begin{theorem}
\label{npnp}
Satisfiability and poor man's satisfiability with respect to the
frame  
\setlength{\unitlength}{0.5pt}
\begin{picture}(30,30)(18,-10)
\put(30,10){\circle*{5}}
\put(27,6){\vector(-3,-4){9}}
\put(30,6){\vector(0,-1){12}}
\put(33,6){\vector(3,-4){9}}
\put(15,-10){\circle*{5}}
\put(30,-10){\circle*{5}}
\put(45,-10){\circle*{5}}
\end{picture}
are NP-complete.
\end{theorem}

\begin{proof}
Because the frame is finite, both satisfiability problems are in NP.
Thus it suffices to show that poor man's satisfiability with
respect to 
\setlength{\unitlength}{0.5pt}
\begin{picture}(30,30)(18,-10)
\put(30,10){\circle*{5}}
\put(27,6){\vector(-3,-4){9}}
\put(30,6){\vector(0,-1){12}}
\put(33,6){\vector(3,-4){9}}
\put(15,-10){\circle*{5}}
\put(30,-10){\circle*{5}}
\put(45,-10){\circle*{5}}
\end{picture}
is NP-hard.

Since we are working with a fragment of propositional modal logic,
it is extremely tempting to try to reduce from
an NP-complete propositional satisfiability problem.
However, because poor man's logics contain only a
fragment of propositional logic,
these logics don't behave like propositional logic at all.
Because of this, 
propositional satisfiability problems are not the best choice of
problems to reduce from. In fact, they are particularly confusing.

It turns out that it is much easier to reduce a partitioning problem
to our poor man's satisfiability problem.
We will reduce from  the following well-known NP-complete problem.

\begin{quote}
GRAPH 3-COLORABILITY:
Given an undirected graph $G$, can you color every vertex
of the graph using only three colors in such a way that
vertices connected by an edge have different colors?
\end{quote}

Suppose $G = (V,E)$ where $V = \{1,2,\ldots, n\}$. 
We introduce a propositional variable $p_e$ for every
edge $e$. The three leaves of 
\setlength{\unitlength}{0.5pt}
\begin{picture}(30,30)(18,-10)
\put(30,10){\circle*{5}}
\put(27,6){\vector(-3,-4){9}}
\put(30,6){\vector(0,-1){12}}
\put(33,6){\vector(3,-4){9}}
\put(15,-10){\circle*{5}}
\put(30,-10){\circle*{5}}
\put(45,-10){\circle*{5}}
\end{picture}
will correspond to the three
colors.
To ensure that adjacent vertices in the graph end up in different
leaves, 
we will make sure that the smaller endpoint of $e$ satisfies
$p_e$ and that the larger endpoint of $e$ satisfies $\ol{p_e}$.

The requirements for vertex $i$ are given by the following formula:
\[\psi_i = \bigwedge \{p_e \mbar e = \{i,j\} \mbox { and } i < j \} \wedge
\bigwedge \{\ol{p_e} \mbar e = \{i,j\} \mbox { and } i > j \}.\]

Define $f(G) = \bigwedge_{i = 1}^n \Diamond \psi_i$.

$f$ is clearly computable in polynomial time. To show that $f$ is
indeed a reduction from
GRAPH 3-COLORABILITY to poor man's satisfiability
with respect to 
\setlength{\unitlength}{0.5pt}
\begin{picture}(30,30)(18,-10)
\put(30,10){\circle*{5}}
\put(27,6){\vector(-3,-4){9}}
\put(30,6){\vector(0,-1){12}}
\put(33,6){\vector(3,-4){9}}
\put(15,-10){\circle*{5}}
\put(30,-10){\circle*{5}}
\put(45,-10){\circle*{5}}
\end{picture}
, first note that
it is easy to see that
for every set $V' \subseteq V$, the following holds: 
$\bigwedge_{i \in V'} \psi_i$ is satisfiable if and only if no two vertices
in $V'$ are connected by an edge.  

It follows that 
$f(G) = \bigwedge_{i = 1}^n \Diamond \psi_i$ is satisfiable on
\setlength{\unitlength}{0.5pt}
\begin{picture}(30,30)(18,-10)
\put(30,10){\circle*{5}}
\put(27,6){\vector(-3,-4){9}}
\put(30,6){\vector(0,-1){12}}
\put(33,6){\vector(3,-4){9}}
\put(15,-10){\circle*{5}}
\put(30,-10){\circle*{5}}
\put(45,-10){\circle*{5}}
\end{picture}
if and only if there exist sets of vertices
$V_1, V_2$, and $V_3$ such that $V = V_1 \cup V_2 \cup V_3$ and 
$\bigwedge_{i \in V_j} \psi_i$ is satisfiable for $j \in \{1, 2, 3\}$.
This holds if and only 
if there exist sets of vertices
$V_1, V_2$, and $V_3$ such that $V = V_1 \cup V_2 \cup V_3$ and 
no two vertices in $V_j$ are adjacent for $j \in \{1, 2, 3\}$,
which is the case if and only if $G$ is 3-colorable. 
(We obtain a coloring by coloring each vertex $v$ by the smallest $j$
such that $v \in V_j$.)
\end{proof}

\section{Poor Man's Versions of PSPACE-complete Satisfiability Problems}

It is well-known that the satisfiability problems
for many modal logics including 
$\K$ are PSPACE-complete~\cite{ladner}.
We also know that poor man's satisfiability for $\K$ is
coNP-complete~\cite{ss,dhlsnn}.
That is, in that particular case the complexity of the satisfiability
problem drops from PSPACE-complete to coNP-complete. Is this the general
pattern? We will show that this is not the case.
We will give an example of a logic where the complexity
of the satisfiability problem drops from PSPACE-complete all the
way down to P, and another example in which
the complexity of both the satisfiability and the poor man's
satisfiability problems are PSPACE-complete.
Both examples are really close to $\K$; they are satisfiability
with respect to $\Kgeqone$ and $\Kleqtwo$, respectively.

We will first consider $\Kgeqone$.
This logic is very close to $\K$ and
it should come as no surprise that the complexity of $\Kgeqone$
satisfiability and $\K$ satisfiability are the same.
It may come as a surprise
to learn that poor man's satisfiability
with respect to $\Kgeqone$ is in P. 
It is easy to show that poor man's satisfiability with respect to
$\Kgeqone$ is in coNP,
because the following function $f$ reduces 
the poor man's satisfiability problem with respect to
$\Kgeqone$ to the
poor man's satisfiability problem for $\K$.
\[f(\phi) = \phi \wedge \bigwedge_{i = 0}^{\smd(\phi)} \Box^i \Diamond q,\]
where $q$ is a propositional variable not in $\phi$.
The formula ensures that every world in the relevant part of the
$\K$ frame has at least one successor.

It is very surprising that poor man's satisfiability
with respect to $\Kgeqone$ is in P, because the  relevant
part of the $\Kgeqone$ frame 
may require an exponential number
of worlds to satisfy a formula in poor man's language.
For example, consider the following formula:

\[\Diamond\Box\Box p_1 \wedge \Diamond\Box\Box \ol{p_1} \wedge
\Box(\Diamond\Box p_2 \wedge \Diamond\Box \ol{p_2}) \wedge
\Box\Box(\Diamond p_3 \wedge \Diamond \ol{p_3}).\]

If this formula is satisfiable in world $w$, then for every assignment
to $p_1,p_2,$ and $p_3$, there exists a world reachable in
three steps from $w$
that satisfies that assignment.

In its general version, the formula becomes
\[\phiexp = \bigwedge_{i = 1}^n\Box^{i-1}(\Diamond\Box^{n-i} p_i \wedge
\Diamond\Box^{n-i} \ol{p_i}).\]

The formula is of length polynomial in $n$ and forces the
relevant part of the model to be of exponential size.

Now that we have seen how surprising it is that
poor man's satisfiability with respect to $\Kgeqone$ is in P,
let's prove it.

\begin{theorem}
\label{pspacep}
Satisfiability with respect to $\Kgeqone$ is PSPACE-complete and
poor man's satisfiability 
with respect to $\Kgeqone$ is in P.
\end{theorem}

\begin{proof}
The proof that satisfiability with respect to $\Kgeqone$ is PSPACE-complete is
very close to the proof that $\K$ satisfiability is
PSPACE-complete~\cite{ladner} and therefore omitted.

For the poor man's satisfiability problem,
note that a simplified version of Ladner's PSPACE upper
bound construction for $\K$ can be used to
show the following.

Let $\phi = \Box \psi_1 \wedge \cdots \wedge  \Box \psi_k \wedge
\Diamond \xi_1 \wedge \cdots \wedge  \Diamond \xi_m \wedge
\ell_1 \wedge \cdots \wedge \ell_s$, where the $\ell_i$s are 
literals. $\phi$ is $\Kgeqone$ satisfiable if and only if
\begin{enumerate}
\item
$\ell_1 \wedge \cdots \wedge \ell_s$ is satisfiable, 
\item for all $j$,
$\psi_1 \wedge \cdots \wedge \psi_k \wedge \xi_j$ is $\Kgeqone$ satisfiable, and
\item
$\psi_1 \wedge \cdots \wedge \psi_k$ is $\Kgeqone$ satisfiable.
(only relevant when $m = 0$.)
\end{enumerate}

Note that this algorithm takes exponential time and polynomial space.
Of course, we already know that poor man's satisfiability
with respect to $\Kgeqone$ is in PSPACE, since 
satisfiability with respect to $\Kgeqone$ is in PSPACE.
How can this PSPACE algorithm help to prove that
poor man's satisfiability with respect to  $\Kgeqone$ is in P?

Something really surprising happens here.
We will prove that for every
poor man's formula $\phi$, $\phi$ is $\Kgeqone$ satisfiable
if and only if (the conjunction of) every pair of
(not necessary different) conjuncts of $\phi$ is
$\Kgeqone$ satisfiable. Using dynamic programming, we can compute all pairs 
of subformulas of $\phi$ that are $\Kgeqone$ satisfiable in polynomial time.
This proves the theorem. It remains to show that
for every poor man's formula $\phi$, $\phi$ is $\Kgeqone$ satisfiable
if and only if every pair of conjuncts of $\phi$ is
$\Kgeqone$ satisfiable. 
We will prove this claim
by induction on $\md(\phi)$, the modal depth of $\phi$.
In the proof, we will write ``satisfiable'' for ``satisfiable
with respect to $\Kgeqone$.''

If $\md(\phi) = 0$, $\phi$
is a conjunction of literals. In that case $\phi$ is not satisfiable
if and only if there exist $i$ and $j$ such that $\ell_i = \ol{\ell_j}$.
This immediately implies our claim.

For the induction step, 
suppose $\phi = \Box \psi_1 \wedge \cdots \wedge  \Box \psi_k \wedge
\Diamond \xi_1 \wedge \cdots \wedge  \Diamond \xi_m \wedge
\ell_1 \wedge \cdots \wedge \ell_s$ (where the $\ell_i$s are 
literals), $\md(\phi) \geq 1$, and suppose that our
claim holds for all formulas of modal depth $< \md(\phi)$.
Suppose for a contradiction that $\phi$ is not satisfiable,
though every pair of conjuncts
of $\phi$ is satisfiable. Then, by the Ladner-like construction given
above, we are in one of the following three cases:

\begin{enumerate}
\item
$\ell_1 \wedge \cdots \wedge \ell_s$ is not satisfiable, 
\item for some $j$, 
$\psi_1 \wedge \cdots \wedge \psi_k \wedge \xi_j$ is  not satisfiable, or
\item
$\psi_1 \wedge \cdots \wedge \psi_k$ is not satisfiable.
\end{enumerate}
By induction, it follows immediately that we are in one of the following
four cases:
\begin{enumerate}
\item
There exist $i, i'$ such that $\ell_i \wedge \ell_{i'}$ is not satisfiable, 
\item
there exist $i, i'$ such that $\psi_i \wedge \psi_{i'}$ is not satisfiable, 
\item 
there exist $i, j$ such that
$\psi_i \wedge \xi_{j}$ is not satisfiable, or
\item 
there exists a $j$ such that
$\xi_{j} \wedge \xi_j$ is not satisfiable.
\end{enumerate}
If we are in case 2, 
$\Box \psi_i \wedge \Box \psi_{i'}$ is not satisfiable.
In case 3, $\Box \psi_i \wedge \Diamond \xi_{j}$ is not satisfiable.
In case 4, $\Diamond \xi_{j} \wedge \Diamond \xi_j$ is not satisfiable.
So in each case we have found a pair of conjuncts of $\phi$
that is not satisfiable, which contradicts the assumption.
\end{proof}

\medskip

Why doesn't the same construction work for $\K$? It is easy
enough to come up with a counterexample. For example,
$\{\Box p, \Box \ol{p}, \Diamond q\}$ is not satisfiable,
even though every pair is satisfiable.
The deeper reason is that we have some freedom in 
$\K$ that we don't have in $\Kgeqone$.
Namely, on a $\K$ frame a world can have successors or no 
successors.  This little bit of extra freedom is enough to 
encode coNP in poor man's language.

\medskip

Theorem~\ref{npnp} showed that poor man's satisfiability can be 
as hard as satisfiability for NP-complete logics. 
In light of the fact that poor man's satisfiability for $\K$
is coNP-complete and  poor man's satisfiability
with respect to $\Kgeqone$ is  even in P, you might wonder if the complexity of
PSPACE-complete logics always decreases.

To try to keep the complexity as high as possible, it makes sense
to look at frames in which each world has a restricted number
of successors, as in the construction of Theorem~\ref{npnp}.
Because we want the logic to be PSPACE-complete, we also need to make
sure that the frames can simulate binary trees.  The obvious class of
frames to look at is $\Kleqtwo$ --  
the class of frames in which each world has at most two successors. 
This gives us the desired example.

\begin{theorem}
\label{pspacepspace}
Satisfiability and poor man's satisfiability with respect to
$\Kleqtwo$ are PSPACE-complete.
\end{theorem}

\begin{proof}
Satisfiability with respect to $\Kleqtwo$
is PSPACE-complete by pretty much the same
proof as the PSPACE-completeness proof for $\K$~\cite{ladner}.
To show that the poor man's version
remains PSPACE-complete, first note that a formula is $\Kleqtwo$ satisfiable
if and only if it is satisfiable in the root of a binary tree.
Stockmeyer~\cite{sto} showed that the set of true
quantified 3CNF formulas is PSPACE-complete.
Using padding, it is immediate that the  following variation
of this set is also PSPACE-complete. 

\begin{quote}
QUANTIFIED 3SAT:
Given a quantified Boolean formula
$\exists p_1 \forall p_2 \exists p_3 \cdots \exists p_{n-1} \forall p_n
\phi$,
where $\phi$ is a propositional
formula over $p_1, \ldots, p_n$ in 3CNF
(that is, a formula in conjunctive normal form with
exactly 3 literals per clause),
is the formula true?
\end{quote}

We will reduce QUANTIFIED 3SAT 
to poor man's satisfiability with respect to binary trees.
To simulate the quantifiers, we need to go back to the formula that
forces models to be of exponential size.

\[\phiexp = \bigwedge_{i = 1}^n\Box^{i-1}(\Diamond\Box^{n-i} p_i \wedge
\Diamond\Box^{n-i} \ol{p_i}).\]
$\phiexp$ is clearly satisfiable in the root of a binary tree and
if $\phiexp$ is satisfied in the root of a binary tree,
the worlds of depth $\leq n$ form a complete binary tree
of depth $n$ and every assignment to $p_1,\ldots, p_n$ occurs exactly once
in a world at depth $n$.
We will call the worlds at depth $n$ the assignment-worlds.

The assignment-worlds in a subtree
rooted at a world at distance $i \leq n$ from the
root are constant with respect to the value of  $p_i$.
It follows that
$\exists p_1 \forall p_2 \exists p_3 \cdots \exists p_{n-1} \forall p_n
\phi \in $ QUANTIFIED 3SAT if and only if 
$\phiexp \wedge
(\Diamond \Box)^{n/2} \phi$ is satisfiable
with respect to binary trees.

This proves that satisfiability for $\Kleqtwo$ is PSPACE-hard,
but it does {\em not} prove that the  poor man's version
is PSPACE-hard.
Recall that $\phi$ is in 3CNF and thus
not a poor man's formula.

Below, we will show how to label all assignment-worlds where
$\phi$ does not hold by $f$ (for $\false$). It then suffices to add
the conjunct $(\Diamond \Box)^{n/2} \ol{f}$
to obtain a reduction.

How can we label all assignment-worlds
where $\phi$ does not hold by $f$?
Let $k$ be such that
$\phi = \psi_1 \wedge \psi_2 \wedge \cdots \wedge \psi_k$,
where each $\psi_i$ is the disjunction of exactly 3 literals:
$\psi_i = \ell_{i1} \vee \ell_{i2} \vee \ell_{i3}$.
We assume without loss of generality
that $n$ is even and
that each $\psi_i$ contains 3 different propositional variables.

For every $i$, we will label
all assignment-worlds where $\psi_i$ does not hold 
by $f$. Since 
$\psi_i = \ell_{i1} \vee \ell_{i2} \vee \ell_{i3}$, this implies that
we have to label all assignment-worlds where
$\ol{\ell_{i1}} \wedge \ol{\ell_{i2}} \wedge \ol{\ell_{i3}}$ holds by $f$.
In general, this cannot be done in poor man's logic,
but in this special case we are able to do it,
because the relevant part of the model is completely fixed
by $\phiexp$.

As a warm-up, first consider how you would label all assignment-worlds
where $\ol{p_3}$ holds by $f$.
This is easy; add the conjunct
\[\Box\Box\Diamond\Box^{n-3}(\ol{p_3} \wedge f).\]

You can label all assignment-worlds where
$\ol{p_3}\wedge p_5$ holds as follows:
\[\Box\Box\Diamond\Box\Diamond\Box^{n-5}(\ol{p_3} \wedge  p_5 \wedge f).\]

This can easily be generalized to a labeling for
$\ol{p_3} \wedge p_5 \wedge \ol{p_8}$:
\[\Box\Box\Diamond\Box\Diamond\Box\Box\Diamond\Box^{n-8}
(\ol{p_3} \wedge  p_5 \wedge \ol{p_8} \wedge f).\]

Note that we can write the previous 
formula in the following suggestive way:
\[\Box^{3-1}\Diamond\Box^{5-3-1}\Diamond\Box^{8-5-1}\Diamond\Box^{n-8}
(\ol{p_3} \wedge  p_5 \wedge \ol{p_8} \wedge f).\]

In general, suppose you want to label all assignment-worlds where
$\ell_1 \wedge \ell_2 \wedge \ell_3$ hold by $f$, where $\ell_1$, $\ell_2$,
and $\ell_3$ are literals. Suppose that $\ell_1$, $\ell_2$, and $\ell_3$'s
propositional variables are $p_{a}$, $p_b$, and $p_c$, respectively.
Also suppose that $a < b < c$.
The labeling formula
$\labelfalse(\ell_1 \wedge \ell_2 \wedge \ell_3)$ is defined as
follows.

\[\labelfalse(\ell_1 \wedge \ell_2 \wedge \ell_3) =
\Box^{a-1}\Diamond\Box^{b-a-1}\Diamond\Box^{c-b-1}\Diamond\Box^{n-c}
(\ell_1 \wedge  \ell_2 \wedge \ell_3 \wedge f).\]

If $\labelfalse(\ell_1 \wedge \ell_2 \wedge \ell_3)$
is satisfied in the root of a complete binary tree, then there exist
at least $2^{n-3}$ worlds at depth $n$ such that 
$(\ell_1 \wedge  \ell_2 \wedge \ell_3 \wedge f)$ holds.

If $\phiexp$ is satisfied in the root of a binary tree, then
the worlds of depth $\leq n$ form a complete binary tree
and there are exactly $2^{n-3}$ assignment-worlds such that
$(\ell_1 \wedge  \ell_2 \wedge \ell_3)$ holds.

It follows that if 
$\phiexp$ is satisfied in the root of a binary tree, then
$\labelfalse(\ell_1 \wedge \ell_2 \wedge \ell_3)$
is satisfied in the root if and only if
$f$ holds in every assignment-world where
$(\ell_1 \wedge  \ell_2 \wedge \ell_3)$ holds.

Thus, the following function $g$ is a reduction from QUANTIFIED 3SAT
to poor man's satisfiability
with respect to $\Kleqtwo$.
\[g(\exists p_1 \forall p_2 \exists p_3 \cdots \exists p_{n-1} \forall p_n
\phi) = 
\phiexp \wedge \bigwedge_{i = 1}^k
\labelfalse( \ol{\ell_{i1}} \wedge \ol{\ell_{i2}} \wedge \ol{\ell_{i3}})
\wedge (\Diamond \Box)^{n/2} \ol{f}. \]
\end{proof}

\medskip

Why doesn't the construction of Theorem~\ref{pspacepspace} work for $\K$?
A formula that is satisfiable in a world with exactly two successors is also
satisfiable in a world with more than two successors.
Because of this, the $\labelfalse$ formula will not necessarily label
all assignment-worlds where $\phi$ does not hold by $f$.
For a very simple example, consider the formula

\[\Diamond p \wedge \Diamond \ol{p} \wedge \Diamond (p \wedge f) \wedge
\Diamond (\ol{p} \wedge f) \wedge \Diamond \ol{f}.\]
This formula is not  $\Kleqtwo$ satisfiable, since both the $p$
successor and the
$\ol{p}$ successor are labeled $f$. However, this formula is 
satisfiable in a world with three successors, satisfying
$p \wedge f$, $\ol{p} \wedge f$, and $\ol{f}$, respectively.

\section{\ALEN\ Satisfiability is PSPACE-complete}
In the introduction, we mentioned that poor man's logic is
closely related to certain description logics. Donini et al.~\cite{dlnn}
almost completely characterize the complexity of the most common description
logics. The only language they couldn't completely
characterize is \ALEN. \ALEN\ is \ALE\ (the
poor man's version of multi-modal $\K$) with number restrictions.
Number restrictions are of the form $(\leq\! n)$ and $(\geq\! n)$.
$(\leq\! n)$ is true if and only if a world has $\leq n$ successors
and $(\geq\! n)$ is true if and only if a world has $\geq n$ successors.

In~\cite{dlnn}, it was shown that 
\ALEN\ satisfiability is in PSPACE, assuming that the number restrictions
are given in unary.
Tobies~\cite{tobies} showed that 
\ALEN\ satisfiability remains in PSPACE if the number restrictions are
given in binary.
The best lower bound for \ALEN\ satisfiability was
the coNP lower bound that is immediate from the fact that this is
an extension of \ALE.

We will use Theorem~\ref{pspacepspace} to prove 
PSPACE-hardness for a very restricted version
of \ALEN.

\begin{theorem}
Satisfiability for the poor man's version of $\K$ extended
with the number restriction $(\leq \! 2)$ is PSPACE-hard.
\end{theorem}

\begin{proof}
The reduction from poor man's satisfiability
with respect to $\Kleqtwo$ is obvious.
It suffices to use the number restriction $(\leq\! 2)$ 
to make sure that every world in the relevant part of the model
has at most two successors.
Let $\md(\phi)$ be the modal depth of $\phi$. All worlds that
are of importance to the satisfiability of $\phi$ are at most $\md(\phi)$
steps away from the root. The reduction is as follows:

\[f(\phi) = \phi \wedge \bigwedge_{i = 0}^{\smd(\phi)} \Box^i
(\leq\! 2) \]
\end{proof}

Combining this with the PSPACE upper bound from~\cite{dlnn}
completely characterizes the complexity of \ALEN\ satisfiability.

\begin{corollary}
\ALEN\ satisfiability is  PSPACE-complete.
\end{corollary}

\section{Other Restrictions on the Set of Operators}
\label{s:other}
As mentioned in the introduction, restricting the modal language in
the way that we have, i.e., looking at formulas built from
literals, $\wedge$, $\Box$, and $\Diamond$, was
motivated by the fact that this restriction occurs in
description logics and also by the rather bizarre complexity behavior
of this fragment.

From a more technical point of view however, we might well wonder
what happens to other restrictions on the set of operators allowed.
After all, who is to say which sublanguages will be useful in the future?
Also, we might hope to gain more insight in the sources
of complexity for modal logics by looking at different sublanguages.

For $S \subseteq \allOp$, let $\calL(S)$ denote the modal language
whose formulas are built from an infinite set of 
propositional variables and operators from
$S$.  We will write $\atneg$ for propositional negation,
and $\neg$ for general negation. So, our ``old'' language $\calL$ will
be denoted by $\calL(\{\atneg, \wedge,\vee,\Box,\Diamond\})$,
and poor man's language by $\lpml$.

In this section, 
we will completely characterize the complexity
of $\calL(S)$ satisfiability (with respect to
the class of all frames), for every $S \subseteq \allOp$.

Since there are $2^8$ subsets, this may seem to be a daunting task.
But we will see that there are only four possibilities
for the complexity of these satisfiability problems:
P, NP-complete, coNP-complete, and PSPACE-complete.  Also, there are not many
surprises: Languages
that contain a complete basis for modal logic obviously have
PSPACE-complete satisfiability problems (Theorem~\ref{pspace}), 
languages that contain a complete basis for propositional
logic, but not for modal logic have NP-complete
satisfiability problems (Theorem~\ref{np}), and
poor man's logic is the only constant-free coNP-complete case
(Theorem~\ref{conp}(1)).
All other cases are in P (Theorem~\ref{p},~\ref{covered}),
except for the surprise that
$\calL(\{\wedge, \Box, \Diamond, \false\})$
satisfiability is coNP-complete (Theorem~\ref{conp}(2))
and that
$\calL(\{\wedge, \vee, \Box, \Diamond, \false\})$ 
satisfiability is PSPACE-complete (Theorem~\ref{pspacetwo}).

\begin{theorem}
\label{pspace}
If $S \subseteq \allOp$ contains $\{\neg, \wedge, \Box\}$,
$\{\neg, \wedge, \Diamond\}$,
$\{\neg, \vee, \Box\}$,
$\{\neg, \vee, \Diamond\}$, or
$\{\atneg, \wedge, \vee, \Box, \Diamond\}$,
then $\calL(S)$ satisfiability is PSPACE-complete.
\end{theorem}

\begin{proof} It is easy to see that these five sets are
exactly the minimal complete bases for modal logic.
All these satisfiability problems are polynomial-time equivalent,
and their PSPACE-completeness follows immediately from Ladner~\cite{ladner}.
\end{proof}

\begin{theorem}
\label{np}
\begin{enumerate}
\item
If $S \subseteq \{\neg, \atneg, \wedge, \vee, \true, \false\}$ and
$S$ contains $\{\neg, \vee\}$ or $\{\neg, \wedge\}$, then
$\calL(S)$ satisfiability is NP-complete.
\item
If $S$ contains $\{\atneg, \wedge, \vee\}$ and
$S$ is a subset of 
$\{\atneg, \wedge, \vee, \Box, \true, \false\}$ or of
$\{\atneg, \wedge, \vee, \Diamond, \true, \false\}$, then
$\calL(S)$ satisfiability is NP-complete.
\end{enumerate}
\end{theorem}

\begin{proof} First note that all these cases are
clearly NP-hard, because they contain propositional satisfiability.
It remains to show that $\calL(S)$ satisfiability is in NP
if $S$ =
$\{\neg, \atneg, \wedge, \vee, \true, \false\}$, 
$\{\atneg, \wedge, \vee, \Box, \true, \false\}$, or
$\{\atneg, \wedge, \vee, \Diamond, \true, \false\}$.

The first case is exactly propositional satisfiability
and thus in NP.

For the second case,  an
$\calL(\{\atneg, \wedge, \vee, \Box, \true, \false\})$ formula is
satisfiable if and only if 
substituting $\true$ for every outermost $\Box \psi$ subformula
gives a propositionally satisfiable formula.

For the last case, the following NP algorithm decides
$\calL(\{\atneg, \wedge, \vee, \Diamond, \true, \false\})$ satisfiability.
Given a formula $\phi$, guess a valuation on all the subformulas of
$\phi$ and accept if and only if this valuation makes
$\phi$ true, the valuation is propositionally consistent, and
for all $\Diamond \psi$ that are set to true, 
$\psi$ is satisfiable.  It is crucial that we do not have to verify
anything if $\Diamond \psi$ is set to false in the valuation,
because $\Diamond \psi$ can only occur positively.
\end{proof}

\begin{theorem}
\label{conp}
\begin{enumerate}
\item
If $\{\atneg, \wedge, \Box, \Diamond\} \subseteq S
\subseteq \{\atneg, \wedge, \Box, \Diamond, \true, \false\}$,
then $\calL(S)$ satisfiability is coNP-complete.
\item
If $\{\wedge, \Box, \Diamond, \false\} \subseteq S
\subseteq \{\atneg, \wedge, \Box, \Diamond, \true, \false\}$,
then $\calL(S)$ satisfiability is coNP-complete.
\end{enumerate}
\end{theorem}

\begin{proof}
As mentioned in the introduction, the
$\calL(\{\atneg, \wedge, \Box, \Diamond, \true, \false\})$
case follows from~\cite{ss,dhlsnn}.

It is easy to see that
$\calL(\{\atneg, \wedge, \Box, \Diamond, \true, \false\})$ satisfiability
with respect to any class of frames ${\cal F}$ is polynomial-time
reducible to $\calL(\{\atneg, \wedge, \Box, \Diamond\})$ satisfiability
with respect to the same class of frames:  Introduce two new variables
$t$ and $f$ to simulate $\true$ and $\false$, respectively,
and define reduction $g$ as follows.

\[g(\phi) = \phi[\true := t, \false := f]
\wedge \bigwedge_{i=0}^{\smd(\phi)}\Box^{i}(t \wedge \ol{f}).\]
This completes the proof of part (1), i.e., poor man's
logic with or without constants.

It remains to show that
$\calL(\{\wedge, \Box, \Diamond, \false\})$ satisfiability is
coNP-complete.
It follows from careful inspection of the proof
of~\cite[Theorem 3.3]{dhlsnn} that satisfiability for variable-free 
$\calL(\{\wedge, \Box, \Diamond, \true, \false\})$ formulas
is coNP-hard.

Let $\phi$ be an $\calL(\{\wedge, \Box, \Diamond, \true, \false\})$
formula.  It is easy to see that $\phi$ is satisfiable if and only if 
the $\calL(\{\wedge, \Box, \Diamond,\false\})$ formula
$\phi[\true := t] \wedge \bigwedge_{i = 0}^{\smd(\phi)} \Box^i t$
is satisfiable.
\end{proof}

\begin{theorem}
\label{p}
If $S$ is a subset of 
$\{\neg, \atneg, \Box, \Diamond, \true, \false\}$,
$\{\atneg, \vee, \Box, \Diamond, \true, \false\}$,
$\{\atneg, \wedge, \Box, \true, \false\}$,
$\{\atneg, \wedge, \Diamond, \true, \false\}$,
$\{\wedge, \vee, \Box, \true, \false\}$,
$\{\wedge, \vee, \Diamond, \true, \false\}$, or
$\{\wedge, \vee, \Box, \Diamond, \true\}$,
then
$\calL(S)$ satisfiability is in P.
\end{theorem}

\begin{proof}~\begin{enumerate}
\item
Every $\calL(\{\neg, \atneg, \Box, \Diamond, \true, \false\})$ formula
can in polynomial time be transformed into an equivalent
$\calL(\{\atneg, \Box, \Diamond, \true, \false\})$ formula, by moving the
negations inward. Since all operators are unary, every 
$\calL(\{\atneg, \Box, \Diamond, \true, \false\})$ formula is
of the form $\{\Box,\Diamond\}^* a$, where
$a$ is a literal or a constant.
It is easy to see that the unsatisfiable formulas of this
form are exactly the formulas of the form $\Diamond^* \false$.

\item
$\calL(\{\atneg, \vee, \Box, \Diamond, \true, \false\})$ 
formulas are of the form
$\Box \psi_1 \vee \cdots \vee  \Box \psi_k \vee
\Diamond \xi_1 \vee \cdots \vee  \Diamond \xi_m \vee
a_1 \vee \cdots \vee a_s$, where the $a_i$s are 
literals or constants, and the $\psi_i$s and $\xi_i$s
are $\calL(\{\atneg, \vee, \Box, \Diamond, \true, \false\})$
formulas.

If $k > 0$ or if one of the $a_i$s is a literal
or $\true$, this formula is satisfiable.
Otherwise, the formula is satisfiable if and only if
one of the $\xi_i$s is satisfiable.
This gives a recursive polynomial-time algorithm. 

\item
$\calL(\{\atneg, \wedge, \Box, \true, \false\})$ formulas are of the form
$\Box \psi_1 \wedge \cdots \wedge  \Box \psi_k \wedge
a_1 \wedge \cdots \wedge a_s$, where the $a_i$s are 
literals or constants. This formula is satisfiable if and only
if $a_1 \wedge \cdots \wedge a_s$ is satisfiable.

\item
$\calL(\{\atneg, \wedge, \Diamond, \true, \false\})$
formulas are of the form:
$\Diamond \xi_1 \wedge \cdots \wedge  \Diamond \xi_m \wedge
a_1 \wedge \cdots \wedge a_s$, where the $a_i$s are 
literals or constants. This formula is satisfiable if and only if
$a_1 \wedge \cdots \wedge a_s$ is satisfiable and
for every $i$, $\xi_i$ is satisfiable.
This is gives a recursive polynomial-time algorithm. 

\item 
An $\calL(\{\wedge, \vee, \Box, \true, \false\})$ formula is satisfiable
if and only if substituting $\true$ for every propositional variable
and for every outermost $\Box$ subformula makes the formula true.

\item
Satisfiability for $\calL(\{\wedge, \vee, \Diamond, \true, \false\})$ formulas
can be recursively computed as follows.
Replace every outermost 
$\Diamond \psi$ subformula by $\true$ if $\psi$ is satisfiable
and by $\false$ if $\psi$ is not satisfiable and
replace  all propositional variables by $\true$. $\phi$ is satisfiable if and
only if the resulting propositional sentence evaluates to true.

\item
Every $\calL(\{\wedge, \vee, \Box, \Diamond, \true\})$ formula
is satisfiable.
\end{enumerate}
\end{proof}

Note that it follows from Theorem~\ref{conp}(2) and Theorem~\ref{p}(7)
that adding the constant $\false$ can increase the complexity.
It follows from the next theorem that the complexity
can increase from P to PSPACE-complete.

\begin{theorem}
\label{pspacetwo}
If $S \subseteq \allOp$ contains
$\{\wedge, \vee, \Box, \Diamond, \false\}$, then
$\calL(S)$ satisfiability is PSPACE-complete.
\end{theorem}
We will in fact prove the following theorem.
\begin{theorem}
\label{novar}
Satisfiability for modal formulas without literals
(but with constants $\true$ and $\false$) is PSPACE-complete.
\end{theorem}

This almost immediately implies Theorem~\ref{pspacetwo}.
For let $\phi$ be a formula without propositional variables.
Let $\phi'$ be the formula that results from $\phi$
by bringing $\phi$ into a negation-less normal form.
Then $\phi$ is satisfiable if and only if
$\phi'[\true := t] \wedge \bigwedge_{i=0}^{\smd(\phi)}\Box^{i}t$
is satisfiable. This proves PSPACE-hardness.  The corresponding
upper bound again follows from Ladner~\cite{ladner}.

\medskip

\begin{proof} 
We will use the following theorem.

\begin{theorem}[Halpern~\cite{halpern}]
\label{onevar}
Satisfiability for modal formulas with one propositional variable
is PSPACE-complete. 
\end{theorem}

We will reduce
satisfiability for modal formulas with {\em one} propositional variable to
satisfiability for modal formulas with {\em zero} propositional variables.
The reduction has the same flavor as the proof of
Theorem~\ref{onevar} from~\cite{halpern},
in that we will encode the truth of the 
propositional variable by the frame.

Let $\phi$ be a modal formula with one propositional variable.
We assume that $\phi$ is built from the sole propositional variable
$p$, and operations $\{\neg, \wedge, \Box\}$.

The main idea of the reduction is the following.
Suppose $\phi$ is satisfiable. Then $\phi$ is satisfiable on a model
$M = \pair{W,R,\pi}$ in which every path has length $\leq \md(\phi)$.
We extend $M$ in such a way that the assignment to
$p$ is encoded in the frame.
Define $M' = \pair{W',R',\pi'}$ as follows.
\begin{itemize}
\item
$W' = W \cup \{w_1, w_2, \ldots , w_{\smd(\phi)+1}\}$.
\item
$R' = R \cup \{\pair{w_i,w_{i+1}} \mbar 1 \leq i \leq \md(\phi)\} \cup
\{\pair{w,w_1} \mbar w \in W \mbox{ and } M,w \models p\}$. 
\item
$\pi'$ is irrelevant, since there are no propositional variables.
\end{itemize}
Thus, $M'$ contains all the information
of $M$, and there is a maximal path of length $\md(\phi) + 1$ from a world
$w \in W$ in $M'$ if and only if $M,w \models p$.
We will simulate $p$ by formula $\Diamond^{\smd(\phi)+1}\Box\false$.

Suppose $M,w \models \Box \psi$ for an arbitrary subformula
$\Box \psi$ of $\phi$.
In the reduction, we need to make sure that we do not
force $\psi$ to be true on the new world $w_1$.
So, we will enforce that $\psi$ holds in all successor worlds that do {\em not}
have a maximal path of length $\md(\phi)$. Let $f(\phi) = 
f_{\smd(\phi)}(\phi)$, where $f_{k}(\phi)$ is defined inductively
as follows.

\begin{itemize}
\item $f_k(p) = \Diamond^{k+1} \Box \false$
\item $f_k(\neg \psi) = \neg f_k(\psi)$
\item $f_k(\psi \wedge \xi) = f_k(\psi) \wedge  f_k(\xi)$
\item $f_k(\Box \psi) = \Box(\Diamond^{k}\Box \false \vee f_k(\psi))$
\end{itemize}

It is straightforward to show that for all $w \in W$ and
all formulas $\psi$ with $p$ as only propositional variable and
such that $\md(\psi) \leq \md(\phi)$,
$M,w \models \psi$ if and only if $M',w \models f_{\smd(\phi)}(\psi)$.
This implies that if $\phi$ is satisfiable, then
$f(\phi)$ is satisfiable.

For the converse, suppose that $f(\phi)$ is satisfiable.
Let $M' = \pair{W',R',\pi'}$ be an acyclic model and $w_0 \in W'$ such
that $M',w_0 \models f(\phi)$. Define $M = \pair{W,R,\pi}$ as follows.
\begin{itemize}
\item $W = (W' \setminus \{w \in W' \mbar M',w \models
\Diamond^{\smd(\phi)}\Box\false\}) \cup \{w_0\}$.
\item $R = R' \cap (W \times W)$.
\item $\pi(p) = \{w \in W \mbar M',w \models
\Diamond^{\smd(\phi)+1} \Box \false\}$.
\end{itemize}

Again, it is easy to show that for all $w \in W$ and
all formulas $\psi$ with $p$ as only propositional variable and
such that $\md(\psi) \leq \md(\phi)$,
$M,w \models \psi$ if and only if $M',w \models f_{\smd(\phi)}(\psi)$.

It follows that $\phi$ is satisfiable if and only if $f(\phi)$ is satisfiable.
This proves the theorem, since $f$ is clearly computable in polynomial
time.
\end{proof}

\medskip

It remains to show that we covered all cases.
\begin{theorem}
\label{covered}
If $S \subseteq \{\neg, \atneg, \wedge, \vee, \Box, \Diamond, 
\true, \false\}$,
then one of Theorems \ref{pspace}, \ref{np}, \ref{conp}, \ref{p}, or
\ref{pspacetwo} applies.
\end{theorem}
\begin{proof}
First suppose that $\neg \in S$. If Theorem~\ref{pspace} does
not apply, then $S \subseteq \{\neg, \atneg,\Box, \Diamond, \true, \false\}$
or $S \subseteq \{\neg, \atneg, \wedge, \vee,\true, \false\}$. In the first
case, Theorem~\ref{p}(1) applies. For the second case, either
Theorem~\ref{np} applies or
$S \subseteq \{\neg, \atneg, \true, \false\}$, in which case
Theorem~\ref{p}(1) applies. 

Next suppose that $\neg \not \in S$ and $\atneg \in S$.
If Theorem~\ref{p} does not apply, then
$S$ is not a subset of 
$\{\atneg, \vee, \Box, \Diamond, \true, \false\}$,
$\{\atneg, \wedge, \Box, \true, \false\}$, or
$\{\atneg, \wedge, \Diamond, \true, \false\}$.
This implies that
$S$ contains $\{\atneg,\wedge,\vee\}$ or $\{\atneg, \wedge, \Box, \Diamond\}$.
In the first case, Theorem~\ref{pspace}
applies or Theorem~\ref{np}(2) applies. In the second case,
Theorem~\ref{pspace} applies or Theorem~\ref{conp}(1) applies.

Finally, suppose that $\neg \not \in S$ and $\atneg \not \in S$.
That is, we are in a fragment of positive modal logic
(see for example Dunn~\cite{dunn}).
If Theorem~\ref{p} does not apply, then 
$\false \in S$, since
$S \not \subseteq \{\wedge, \vee, \Box, \Diamond, \true\}$.
Also, $\wedge \in S$, since
$S \not \subseteq \{\atneg, \vee, \Box, \Diamond, \true, \false\}$.
If Theorem~\ref{p} does not apply, then 
$S \setminus \{\wedge, \false\}$ is not a subset of
$\{\vee, \Box, \true\}$ or of $\{\vee, \Diamond, \true\}$. 
It follows that $S$ contains $\{\wedge,\Box,\Diamond,\false\}$ and 
Theorem~\ref{conp}(2) or Theorem~\ref{pspacetwo} applies.
\end{proof}

{\samepage
\begin{center}
{\bf Acknowledgments}
\end{center}
\nopagebreak
\indent
I would like to thank Johan van Benthem for suggesting this topic,
Michael Bauland, Johan van Benthem, Hans de Nivelle, Maarten de Rijke,
Henning Schnoor, and Ilka Schnoor
for helpful conversations and suggestions, and
Stephan Tobies and 
the anonymous referees for useful comments and suggestions.
}

\end{document}